\documentclass{article}
\usepackage[cp1251]{inputenc}
\usepackage[english]{babel}
\usepackage{anysize}
\usepackage{amsmath}
\usepackage{amsfonts}
\usepackage{lscape}
\usepackage[pdftex]{graphicx}
\graphicspath{{images/}}

\begin{document}
\title{General equation for directed Electromagnetic Pulse Propagation in 1D metamaterial: Projecting Operators Method }
\author{\vspace{6pt} Dmitrii Ampilogov, Sergey Leble \\  \small 1. Immanuele Kant Baltic Federal University,  Al. Nevsky st 14, Kaliningrad,  Russia}
\maketitle
\begin{abstract}
We consider a  boundary problem for 1D electrodynamics modeling of a pulse propagation in a
 metamaterial medium.
We build and apply  projecting operators to a Maxwell system in time domain that allows to split the linear propagation
problem to directed waves   for a  material relations with general dispersion. Matrix elements  of the projectors act as convolution integral operators.  For a weak nonlinearity we generalize the linear results still for arbitrary dispersion and  derive the system of interacting right/left waves with combined (hybrid)  amplitudes. The result is specified for the popular metamaterial model with Drude formula for both permittivity and permeability coefficients.  We also discuss and investigate stationary solutions of the system related to some boundary regimes.
\end{abstract}

\section*{Introduction}

\subsection*{Researches on metamaterials}
 The  researches  on metamaterials go up to  Bengali polymath, physicist J. C. Bose publication in 1898 \cite{Bose}. He studied  the  of electric waves polarisation  plane rotation by a special  structure created by him.  W. E. Kock suggested to use  a mixture of metal spheres as a refractive material \cite{Kock1}. He   coined to such materials the term artificial dielectric which has been later used in  a literature on microwaves \cite{Brown}.   

One of most intriguing problems for researchers relate to artificial materials named as  metamaterials with simultaneously negative dielectric permittivity and magnetic permeability. The ideas connected with negative refraction index appeared in 1940-1950s.  L.I. Mandelshtam described negative index refraction and backward propagation of waves \cite{Mandel}. He  based on works of Lamb \cite{Lamb}, Schuster \cite{Shuster} and Pocklington \cite{Pock}.  Malyuzhinets described backward-wave transmission lines in 1951 \cite{Malyu}. In 1957 Sivukhin \cite{Sivu} predicted the possibility of existing of structures with negative $\varepsilon$ and $\mu$. Pafomov \cite{Pafomov} discussed the Vavilov-Cherenkov radiation in such type materials. Only in 1968 Victor Veselago \cite{Veselago} writes about the general electrodynamic properties of metamaterials. 

In 2000 David Smith and his group created such type of structures \cite{Smit}.   Structures with simultaneously negative dielectric permittivity and magnetic permeability has been called by many names: Veselago media, negative-index media, negative-refraction media, backward wave media, double-negative media, negative phase-velocity media, and even left-handed media (LHM) \cite{Sihlova}.  
Since the discovery of materials with negative refractive index, it has been possible to built new devices that use metamaterials ability to control the path of electromagnetic energy. 
The applications for metamaterials are broad and varied from the celebrated electromagnetic cloaking \cite{Shelby}, to new imaging capabilities \cite{Lipworth}. A practical implementation of a subwavelength resonator  presented in \cite{Sanada}. The idea for dispersion compensation in transmission lines using negative-refractive media (NRM)
 was described in \cite{Cheng}. An  interaction of ultrashort pulses with ordinary materials is well understood in nonlinear optics \cite{Wayne} and extended for metamaterals in  \cite{Wen}.

To obtain negative values of the constitutive parameters $\varepsilon$ and $\mu$, metamaterials must be dispersive, i.e., their permittivity and permeability must be frequency dependent, otherwise they would not be causal \cite{Ziol}. As it is shown \cite{Mil} if we have frequency dispersion, full energy density of electromagnetic  field will be:
 \begin{equation}
W = \frac{\partial (\omega \varepsilon(\omega)) }{\partial \omega} E^2 +\frac{\partial (\omega \mu(\omega)) }{\partial \omega} H^2 \label{energy}
 \end{equation}
$W > 0$ if:
$$\frac{\partial (\omega \epsilon(\omega)) }{\partial \omega} >0 , \,\,\, \frac{\partial  (\omega \mu(\omega))  }{\partial \omega} > 0  $$
That's not contradict with  simultaneously negative $\epsilon < 0$ and $\mu < 0$. \cite{Veselago}. 

The materials with typical plasma dispersion (Drude formula, of Lorentz origin) for both $\epsilon(\omega)$ and $\mu(\omega)$   is often  discussed \cite{Ziol2}:
\begin{equation}
\chi = \frac{\omega^2_p \chi_a + i \omega_p \chi_\beta \omega - \chi_\gamma \omega^2}{\omega^2_0 + i \omega \Gamma - \omega^2}
\end{equation}
$\Gamma$ is plasma collision frequency, used as a parameter for a metamaterial. 
There would be independent models for the permittivity and permeability
\begin{equation}
\epsilon (\omega) =  (1+\chi_e), \mu (\omega) =  (1+ \chi_m). \label{mode(w)}
\end{equation} 

The real part of this permittivity is  negative for all $\omega < \sqrt{\omega_p^2 + \Gamma^2 }$. The Drude model is obtained with the fakir's bed medium \cite{Rotman}. 
A simple  example of   metamaterial is the fiber array \cite{Pendry}.

\subsection*{Projecting operators approach}

In our work we use the projecting operators approach, originated from \cite{Leble}. That's a general tool of theoretical physics  to split evolution system to a set of equations of the first order in time that naturally include unidirectional equations corresponding to elementary roots of dispersion  equation. It is based on a complete set of projecting operators, each for a dispersion relation root that fixes the corresponding subspace of a linearized fundamental system such as Maxwell equations. The method, compared to one used in \cite{Fle, New, Kin},  allows us to combine \textit{equations} of the complex basic system in algorithmic way with dispersion, dissipation and, after some development, a nonlinearity taken into account and also, introduces combined (hybrid) fields as basic modes. It therefore allows us to effectively formulate a corresponding mathematical problem: initial or boundary conditions in an appropriate physical language in mathematically correct form.

A part of this method contain a transition to new variables, e.g. of the form
\begin{equation*}
\psi^{\pm}=\varepsilon{\frac{1}{2}}E_{i} \pm  \mu{\frac{1}{2}}H_{j},
\end{equation*}
as did Fleck \cite{Fle}, Kinsler \cite{New, Kin}  and Amiranashvili \cite{Amir} in their works.  This part is in a sense similar to the
projection operator method,  use of which we demonstrate here.

\subsection*{Aim and scope}

In this paper we'll apply mentioned method of projecting operators to the problem of wave propagation in 1D-metamaterial with dispersion of both $\varepsilon$ and $\mu$.  The main aim of the work is very similar to the recent \cite{Kin} and \cite{KuLe} we do want to derive an evolution  equation for the mentioned conditions with the minimal simplifications. The methodical  differences and results are highlighted and discussed.

The article consists of introduction, three sections and conclusion.  

In introduction the actuality of problem and basic ideas of projection method are shown. 

In  section 1 we state the problem. We also show, how the material relations change  while dispersion account.   

The section 2 is devoteto projecting operators construction in $\omega$ and $t$ representations (domains). 

The section three realizes the main task of the solutions space separation. 

In the fourth section we account nonlinearity and realize the important example of the Kerr one, deriving the system of the directed waves interaction. 

The numbers 5 and 6 includes realization of the program for Drude dispersion (5) and Kerr nonlinearity (6) model and hence finalize the main result of the paper:
 the directed waves interaction system for this model, popular in metamaterials investigations. The Sec. 6 includes also the subsection about stationary solutions that show difference between conventional and Veselago materials.

\section{Statement of problem}

\subsection{Maxwell's equations. Operators of dielectric permittivity and magnetic permeability}
Our starting point is the Maxwell equations for a simple case of linear isotropic but dispersive dielectric media, in the SI unit system:
\begin{eqnarray}
  \text{div} \vec{D} (\vec{r},t) &=& 0, \label{eq:Maxwell_D}\\
  \text{div} \vec{B} (\vec{r},t) &=& 0, \label{eq:Maxwell_B} \\
  \text{rot} \vec{E} (\vec{r},t) &=& - \frac{\partial \vec{B} (\vec{r},t)}{\partial t}, \label{eq:Faradey}\\
  \text{rot} \vec{H} (\vec{r},t) &=& \frac{\partial \vec{D} (\vec{r},t)}{\partial t}. \label{eq:Amper}
\end{eqnarray}
We restrict ourselves to a one-dimensional model, similarly to Sch\"afer, Wayne \cite{Wayne} and Kuszner, Leble \cite{KuLe}, where the $x$-axis is chosen as the direction of a pulse propagation. As mentioned authors, we assume $D_x = 0$ and $B_x = 0$, taking into account the only polarization of electromagnetic waves. This allows us to write the Maxwell equations as:
\begin{eqnarray}
 \frac{\partial D_y}{\partial t} &=& -  \frac{\partial H_z}{\partial x}, \label{eq:1DMax2} \\
\nonumber \frac{\partial B_z}{\partial t} &=&  - \frac{\partial E_y}{\partial x}.
\end{eqnarray}
Further indices will be omitted for compactness. We'll introduce four variables $\mathcal{E}$, $\mathcal{B}$, $\mathcal{D}$,  $\mathcal{H}$. They're Fourier images of $E$, $B$,$D$ and $H$ and connected by inverse Fourier transformations:
\begin{eqnarray}
E (x,t) &=& \frac{1}{\sqrt{2 \pi}} \int\limits_{-\infty}^\infty   \mathcal{E}  (x,\omega) \exp(i\omega t) d \omega,  \\
B (x,t) &=& \frac{1}{\sqrt{2 \pi}} \int\limits_{-\infty}^\infty \mathcal{B} (x,\omega) \exp(i\omega t) d \omega , \\
D (x,t) &=& \frac{1}{\sqrt{2 \pi}} \int\limits_{-\infty}^\infty \mathcal{D} (x,\omega) \exp(i\omega t) d \omega \\
H(x,t) &=& \frac{1}{\sqrt{2 \pi }} \int\limits_{-\infty}^\infty \mathcal{H} (x,\omega) \exp(i\omega t) d\omega.
\end{eqnarray}
The domain of Fourier images we'll call $\omega-$representation or a frequency domain. The one of functions $E$, $B$, $D$,$H$ is $t$-representation or a time domain. Linear material equations in $\omega-$ representation we take as:
\begin{equation}
 \mathcal{D} = \varepsilon_0 \varepsilon (\omega) \mathcal{E},\label{eq:mat-D}
\end{equation}
\begin{equation}
\mathcal{B} = \mu_0 \mu (\omega) \mathcal{H}. \label{eq:mat-H}
\end{equation}
Here: \\
$\varepsilon (\omega)$ - dielectric permittivity of medium, $\varepsilon_0$ -dielectric permittivity of vacuum. $\mu (\omega)$ - magnetic permeability of medium and $\mu_0$ - magnetic permeability of vacuum. $\mathcal{B}$ - analogue of function $B$ in $\omega-$representation. For calculation purposes we need to use $t$-representation. In this representation $\varepsilon$ and $\mu$ become integral operators.
Then:
\begin{eqnarray}
D (x,t) &=& \frac{1}{\sqrt{2 \pi}} \int\limits_{-\infty}^\infty \mathcal{D} (x,\omega) \exp(i\omega t) d \omega  = \frac{\varepsilon_0 }{\sqrt{2 \pi}} \int\limits_{-\infty}^\infty  \varepsilon (\omega) \mathcal{E}  (x,\omega) \exp(i\omega t) d \omega, \label{Fu:Ey} \\
B (x,t) &=& \frac{1}{\sqrt{2 \pi}} \int\limits_{-\infty}^\infty \mathcal{B} (x,\omega) \exp(i\omega t) d \omega = \frac{\mu_0}{\sqrt{2 \pi }} \int\limits_{-\infty}^\infty \mu (\omega)\mathcal{H} (x,\omega) \exp(i\omega t) d \omega.  \label{Fu:Bz}
\end{eqnarray}
Plugging
\begin{equation}
\mathcal{E}  (x,\omega)=\frac{1}{\sqrt{2 \pi}} \int\limits_{-\infty}^\infty E(x,s)\exp(-i\omega s)ds \label{PlugE}
\end{equation}
into \eqref{Fu:Ey} we obtain the expression that contains double integral:
\begin{equation}
	D (x,t) = \frac{\varepsilon_0}{2 \pi}  \int\limits_{-\infty}^\infty \varepsilon (\omega) \int\limits_{-\infty}^\infty E(x,s)\exp(-i\omega s)ds \exp(i\omega t) d \omega, \label{int}
\end{equation}
If  $E(x,s)\exp(-i\omega s)$ satisfy basic conditions of Fubini's theorem, we obtain: 
\begin{equation}
	D (x,t) =    \frac{\varepsilon_0}{2 \pi}  \int\limits_{-\infty}^\infty \int\limits_{-\infty}^\infty \varepsilon (\omega)\exp(i\omega (t-s))   d \omega E(x,s)ds= \int\limits_{-\infty}^\infty \tilde{\varepsilon}  (t-s) E(x,s)ds,
\end{equation}
where the kernel
\begin{equation}
	\tilde{\varepsilon}  (t-s)= \frac{\varepsilon_0}{2 \pi} \int\limits_{-\infty}^\infty \varepsilon (\omega)\exp(i\omega (t-s)) d \omega ,
\end{equation}
defines the integral operator of convolution type
\begin{equation}
\widehat{\varepsilon} \psi (x,t) =  \int\limits_{-\infty}^\infty  \tilde{\varepsilon}  (t-s)\psi (x,s)  ds, \label{opEps}
\end{equation}
or
\begin{equation}
	D (x,t) =   \widehat{\varepsilon} E(x,t),
\end{equation}
that expresses the material relation for the definition \eqref{eq:mat-D}.
Doing operations for $E$ and magnetic components of the field, we obtain:
\begin{eqnarray}
E (x,t) =   \int\limits_{-\infty}^\infty \tilde{e}  (t-s) D(x,s)ds,
B (x,t) =  \int\limits_{-\infty}^\infty  \tilde{\mu}(t-s) H (x,s)  ds = \widehat{\mu} H (x,t) ,  \label{opmu}\\
H (x,t) =  \int\limits_{-\infty}^\infty  \tilde{m}(t-s)\mathcal{B} (x,s)  ds = \widehat{\mu}^{-1} B (x,t) , \label{intBs}
\end{eqnarray}
with kernels
\begin{equation}
\begin{aligned}
\tilde{e}  (t-s)= \frac{1}{2 \pi} \int\limits_{-\infty}^\infty \varepsilon^{-1} (\omega)\exp(i\omega (t-s)) d \omega , \\
\tilde{\mu}(t-s) =  \frac{\mu_0}{2 \pi} \int\limits_{-\infty}^\infty \mu (\omega)\exp(i\omega (t-s)) d \omega , \\
\tilde{m}  (t-s)= \frac{1}{2 \pi \mu_0} \int\limits_{-\infty}^\infty \mu^{-1} (\omega)\exp(i\omega (t-s)) d \omega
	\end{aligned}
\end{equation}
The transforms define the fields in time domain, using the conventional continuation of the fields to the half space $t<0$ and
causality condition \cite{Boyd}.

\subsection{Boundary regime problem}

As we see, in time domain, equations for $E$ and $B$     \eqref{eq:1DMax2} may be written in operator form:
\begin{eqnarray}
 \frac{\partial}{\partial t} (\widehat{\varepsilon} E) &=& -  \frac{\partial}{\partial x} (\widehat{\mu}^{-1} B), \label{eq:1DMax3} \\
\nonumber \frac{\partial B}{\partial t} &=&  - \frac{\partial E}{\partial x}.
\end{eqnarray}
Action of operators $\widehat{\varepsilon}$ and $\widehat{\mu}$ was defined by (\ref{opEps}, \ref{intBs}). We also must add the boundary regime conditions:
\begin{equation}
 E (0,t) = j(t), \,\,\, B (0,t) = k(t).
 \label{bc}
\end{equation}

\section{Projecting operators}

Let's use transformations \eqref{Fu:Ey} and \eqref{Fu:Bz}. We substitute them into the system of equation \eqref{eq:1DMax3}:
\begin{equation}
\frac{\varepsilon_0}{\sqrt{2 \pi}} \frac{\partial}{\partial t} \left( \int\limits_{-\infty}^\infty \varepsilon (\omega) \mathcal{E} (x,\omega) \exp(i\omega t) d \omega \right)  = -  \frac{1}{\mu_0 \sqrt{2 \pi}} \frac{\partial}{\partial x} \left(\int\limits_{-\infty}^\infty \frac{\mathcal{B} (x,\omega)}{\mu (\omega)} \exp(i\omega t) d \omega \right) \label{Max2BE}
\end{equation}
The inverse Fourier transformation yields in the first equation of \eqref{eq:1DMax3}
\begin{equation}
\frac{\partial \mathcal{B}}{\partial x} = - i \omega \mu_0 \varepsilon_0 \mu (\omega) \varepsilon (\omega) \mathcal{E}. \label{eq:ByEy}
\end{equation}
The second equation gives similarly:
\begin{equation}
    \frac{\partial \mathcal{E}}{\partial x} = - i \omega \mathcal{B}. \label{eq:ByEy2}
\end{equation}
The notation:
\begin{equation}
\mu_0 \varepsilon_0 \varepsilon (\omega) \mu (\omega) \equiv c^{-2} \varepsilon (\omega) \mu (\omega) \equiv a^2(\omega), \label{def:a(w)}
\end{equation}
where $c$ is the velocity of light in vacuum. By definition:
\begin{equation}
c^2 = \frac{1}{\varepsilon_0 \mu_0}
\end{equation}
Hence the system (\ref{eq:ByEy}, \ref{eq:ByEy2})  simplifies as:
\begin{eqnarray}
  \frac{\partial \mathcal{B}}{\partial x} & = & - i \omega a^2 (\omega) \mathcal{E}, \\
   \frac{\partial \mathcal{E}}{\partial x} & = & - i \omega \mathcal{B}.
\end{eqnarray}
The system \eqref{eq:1DMax3} may be written in matrix form. For this purpose we introduce matrices $\mathcal{L}$ and $\tilde{\Psi}$ as:
\begin{equation}
\tilde{\Psi} = \left(    \begin{array}{c} \mathcal{B} \\ \mathcal{E} \end{array} \right)  \label{koshi-kt}
 \end{equation}
\begin{eqnarray}
\mathcal{L} = \left(\begin{array}{cc}
0 & - i \omega a^2 (\omega)  \\
- i \omega & 0 \end{array} \right) . \label{L-need}
\end{eqnarray}
Hence the matrix form of (\ref{eq:ByEy}, \ref{eq:ByEy2}) is:
\begin{equation}
  \frac{\partial \tilde{\Psi}}{\partial x} = \mathcal{L} \tilde{\Psi}. \label{mxeqw}
 \end{equation}
\eqref{mxeqw} is a system of ordinary differential equations with constant coefficients that have  exponential-type solutions. Following the technique described in \cite{KuLe}, we arrive at $2 \times 2$ eigenvalue problem. Let us look for such matrices $P^{(i)}, i  = \overline{1,2}$ that $P^{(i)} \Psi = \Psi_i$ would be eigenvectors of the evolution matrix \eqref{koshi-kt}. The standard properties of orthogonal projecting operators:
\begin{equation}
\begin{aligned}
P^{(1)} P^{(2)} = 0, i \neq j, \\
P^{(i)} \cdot P^{(i)} = P^{(i)},  \\
P^{(1)} + P^{(2)}  = I,
\end{aligned}
\end{equation}
are implied. It's easy to show, $P^{(i)} $ in $\omega-$representation have the form:
\begin{equation}
P^{(1)} (\omega) =   \frac{1}{2}  \left(\begin{array}{cc}
1 & - a(\omega) \\
- \frac{1}{a(\omega)} & 1 \end{array} \right) \label{P1-wBE},
\end{equation}
\begin{equation}
P^{(2)} (\omega) = \frac{1}{2}   \left(\begin{array}{cc}
1 & a(\omega) \\
\frac{1}{a(\omega)} & 1 \end{array} \right) \label{P2-wBE}.
\end{equation}
Using the inverse Fourier transformation $\widehat{\textbf{P}}^{(i)}  = \mathcal{F} P^{(i)}  \mathcal{F}^{-1}$, where $\mathcal{F}$ - operator of Fourier transformation,  leads to projectors in $t-$representation:
\begin{equation}
\widehat{\textbf{P}}^{(1,2)} (t) =   \frac{1}{2}  \left(\begin{array}{cc}
1 & \widehat{P}^{(1,2)}_{12} \\
\widehat{P}^{(1,2)}_{21} & 1 \end{array} \right), \label{P1-tBE}
\end{equation}
The diagonal elements of projectors are proportional to identity :
\begin{equation}
\begin{aligned}
\widehat{P}^{(1,2)}_{11} \xi(x,t) &=&  \frac{1}{4 \pi} \int\limits_{-\infty}^\infty   \exp(i\omega t-i \omega \tau) \xi(x,\tau) d \omega  =  \frac{1}{2} \int\limits_{-\infty}^\infty  \delta (t-\tau)  \xi(x,\tau) d\tau = \frac{1}{2} \xi (x,t),  \label{P11be}\\
\widehat{P}^{(1,2)}_{22} \eta (t) &=&  \frac{1}{4\pi} \int\limits_{-\infty}^\infty   \exp(i\omega t-i \omega \tau)  \xi (x,\tau) d \omega   = \frac{1}{2} \int\limits_{-\infty}^\infty  \delta (t-\tau) \xi(x,\tau) d\tau = \frac{1}{2} \xi (x,t).
\end{aligned}
\end{equation}
Nondiagonal elements act as   integral operators:
\begin{eqnarray}
\widehat{P}^{(1,2)}_{12} \eta (x,t) =\int\limits_{-\infty}^\infty  p^{(1,2)}_{12}(t,\tau)  \eta (x,\tau) d\tau \label{P12be}, \\
\widehat{P}^{(1,2)}_{21} \xi (x,t) = \int\limits_{-\infty}^\infty  p^{(1,2)}_{21}(t,\tau)  \xi (x,\tau) d\tau , \label{P13be}
\end{eqnarray}
Integral operator kernels of $ \widehat{P}^{(1,2)}_{12}(t,\tau)$ and $\widehat{P}^{(1,2)}_{21}(t,\tau)$ are:
\begin{eqnarray}
   p^{(1,2)}_{12}(t,\tau)  &=&  \mp \frac{1}{4 \pi} \int\limits_{-\infty}^\infty  a(\omega) \exp(i\omega t-i \omega \tau) d \omega ,  \label{P12k} \\
   p^{(1,2)}_{21}(t,\tau)  &=&  \mp \frac{1}{4 \pi} \int\limits_{-\infty}^\infty  \frac{1}{a(\omega)} \exp(i\omega t-i \omega \tau) d \omega ,  \label{P13k}
\end{eqnarray}
So, we can define them via the operator $\widehat{a}$:
\begin{eqnarray}
\widehat{P}^{(1,2)}_{12} \eta (x,t) = \mp \widehat{a}  \eta (x,t)  \label{P12be},\\
\widehat{P}^{(1,2)}_{21} \xi (x,t) =  \mp  \widehat{a}^{-1} \xi (x,t),  \label{P13be}
\end{eqnarray}
where
\begin{equation}\label{hata}
\begin{aligned}
\widehat{a} \eta (x,t) =  \frac{1}{2 \pi} \int\limits_{-\infty}^\infty \left[  \eta (x,\tau) \int\limits_{-\infty}^\infty  a(\omega) \exp(i\omega (t- \tau)) d \omega \right]  d\tau , \\
\widehat{a}^{-1} \xi (x,t) =  \frac{1}{2 \pi} \int\limits_{-\infty}^\infty \left[   \xi (x,\tau)  \int\limits_{-\infty}^\infty  \frac{1}{a(\omega)} \exp(i\omega( t-\tau)) d \omega \right] d\tau .
\end{aligned}
\end{equation}

\section{Separated equations and definition for left and right waves}

We introduce the shorthands:
\begin{equation}
\partial_t \equiv \frac{\partial}{\partial t},\, \partial_x \equiv \frac{\partial}{\partial x}.
\end{equation}
In $t-$representation, matrix equation \eqref{mxeqw} takes the form:
\begin{equation}
\partial_x \Psi = \widehat{L} \Psi, \label{relPsi}
\end{equation}
where
\begin{equation}
\Psi = \left(    \begin{array}{c} B \\ E \end{array} \right),  \label{psit}
 \end{equation}
  \begin{eqnarray}
\widehat{L} = \left(\begin{array}{cc}
0 & - \partial_t  \widehat{a^2}  \\
- \partial_t  & 0 \end{array} \right) . \label{Lt}
\end{eqnarray}

One can  check, that the operator $ \widehat{a^2}$, defined as:
\begin{equation}
 \widehat{a^2} \psi (x,t) = \frac{1}{2\pi} \int\limits_{-\infty}^\infty a^2(\omega) \exp(i\omega t-i \omega \tau)  \psi (x,\tau) d \omega d\tau, \label{def:op:a2}
\end{equation}
acts as square of  $\hat a$, defined by \eqref{hata}.

Making the similar calculations, we can find, that $\widehat{a}^2$ is expressed as product
$$\widehat{a}^2 = \widehat{\varepsilon}\widehat{\mu},$$
that commute
\begin{equation}
\widehat{\varepsilon}  \widehat{\mu}\psi (x,t) = \widehat{\mu} \widehat{\varepsilon} \psi (x,t).
\end{equation}
We note, that this relation is true only if operators $\widehat{\varepsilon}$ and $\widehat{\mu}$ are convolution type integrals. For the further operations we also prove the commutation of operators $\partial_t$ and $\widehat{a}^2$.

 By  acting the operator  $\widehat{\textbf{P}}^{(1)}  $ \eqref{P1-tBE} on the equation \eqref{relPsi} we find:
\begin{equation}
\widehat{\textbf{P}}^{(1)} \partial_x  \Psi  = \widehat{\textbf{P}}^{(1)}  \widehat{L} \tilde{\Psi}.
\end{equation}
We can commute $\widehat{\textbf{P}}^{(1)} $ and $\partial_x$, because projectors doesn't depend on $x$. Using also the proven relations, write
\begin{equation}
\partial_x  \widehat{\textbf{P}}^{(1)} (t)  \Psi = \widehat{\textbf{P}}^{(1)} (t) \widehat{L} \Psi = \widehat{L} \widehat{\textbf{P}}^{(1)}(t)  \Psi. \label{eq:mainpr}
\end{equation}
After substituting $\Psi$ and $\widehat{L}$ (\ref{psit}, \ref{Lt}) and $\widehat{\textbf{P}}^{(1)}$  \eqref{P1-tBE} we find:
\begin{equation}
\partial_x  \left( \begin{array}{c} \frac{1}{2}B + \frac{1}{2} \widehat{a} E \\ \frac{1}{2} \widehat{a}^{-1} B + \frac{1}{2} E \end{array} \right) = \left( \begin{array}{c} - \frac{1}{2} \widehat{a} \partial_t B - \frac{1}{2} \widehat{a}^2 \partial_t E \\ - \frac{1}{2} \partial_t  B - \frac{1}{2} \widehat{a}^{-1} \widehat{a}^2 \partial_t  E \end{array} \right). \label{rel:proj}
\end{equation}
Applying the projection operators to the vector $\Psi$ \eqref{psit}, we can introduce new variables $\Pi$ and $\Lambda$ as:
\begin{equation}
  \Lambda \equiv \frac{1}{2}(B - \widehat{a} E), \label{def:La}
  \end{equation}
  \begin{equation}
  \Pi \equiv \frac{1}{2}(B + \widehat{a} E). \label{def:Pi}
\end{equation}
Those are left and right waves variables. From \eqref{rel:proj} we  get two equations that determine evolution with respect to x of the boundary regime \eqref{bc}:
\begin{equation}
\begin{aligned}
\partial_{x}    \Pi(x,t)   = - \widehat{a} \partial_t \Pi,   \\
 \partial_{x}   \Lambda(x,t) =   \widehat{a}\partial_t  \Lambda.   \label{syslin}
\end{aligned}
\end{equation}

Using relations (\ref{def:Pi}, \ref{def:La}) from \eqref{bc} we derive boundary regime conditions for left and right waves:
\begin{equation} 
\begin{aligned}
\Lambda(0,t) = \frac{1}{2} (B(0,t) - \widehat{a} E(0,t)) = \frac{1}{2} (k(t) - \widehat{a} j(t)) , \\
\Pi(0,t) = \frac{1}{2} (B(0,t) + \widehat{a} E(0,t)) = \frac{1}{2} (k(t) + \widehat{a} j(t)) \label{bcq}
\end{aligned}
\end{equation}
That's  system of operator equations for time-domain dispersion of left and right waves in linear case. \\

\section{Nonlinearity account}

Let's consider a nonlinear problem. We start again from the Maxwell's equations \eqref{eq:1DMax2} with generalized material relations:
\begin{equation}
  \begin{aligned}
D =  \widehat{\varepsilon} E  + P_{NL}, \\
B =  \widehat{\mu} H  + M_{NL},  \label{nlmat}
    \end{aligned}
\end{equation}
$P_{NL}$ - nonlinear part of polarisation ($M_{NL}$ - the one for magnetisation). For our purposes linear parts of polarisation and magnetization have already taken into account.  In time-domain, a closed nonlinear version of \eqref{eq:1DMax3} is:
\begin{eqnarray}
\partial_t (\widehat{\varepsilon} E) + \partial_t P_{NL} &=& -  \partial_x\widehat{\mu}^{-1} B - \partial_x\widehat{\mu}^{-1} M_{NL} , \label{eq:1DMax4} \\
\nonumber \frac{\partial B}{\partial t} &=&  - \frac{\partial E}{\partial x}.
\end{eqnarray}
Action of operator $\widehat{\mu}$ on the first equation of system \eqref{eq:1DMax4} and
using the same notations  $\Psi$ and $\widehat{L}$ from (\ref{psit}, \ref{Lt}) once more, we obtain a nonlinear analogue of the matrix equation \eqref{relPsi}:
\begin{equation}
\partial_x \Psi -  \widehat{L} \Psi  =- \left( \begin{array}{c} \partial_x M_{NL} \\0 \end{array} \right) - \left( \begin{array}{c} \widehat{\mu} \partial_t P_{NL} \\0 \end{array} \right), \label{relPsinl}
\end{equation}
We introduce a vector of nonlinearity:
\begin{equation}
\mathbb{N}(E,B) =  \left(    \begin{array}{c} \partial_x M_{NL} + \partial_t \widehat{\mu} P_{NL} \\0 \end{array} \right)
\end{equation}
Then we get the nonlinear analogue of matrix equation \eqref{relPsi}:
\begin{equation}
\partial_x \Psi   - \widehat{L} \Psi  = -  \mathbb{N}(E,B).  \label{eq:nonl}
\end{equation}
Next,  acting by operators  $\widehat{\textbf{P}}^{(1)} $ \eqref{P1-tBE} and $\widehat{\textbf{P}}^{(2)} $  on the Eq. \eqref{eq:nonl} 
we find:
\begin{equation}\label{main}
\begin{aligned}
\partial_{x}   \Pi   + \widehat{a} \partial_t \Pi  =   \mathbb{N}_1 ( \widehat{a}^{-1} ( \Pi - \Lambda ), \Pi + \Lambda)  , \\  
 \partial_{x}   \Lambda -  \widehat{a} \partial_t  \Lambda = - \mathbb{N}_1 ( \widehat{a}^{-1} ( \Pi - \Lambda ), \Pi + \Lambda),
\end{aligned}
\end{equation}
where 
\begin{equation}
\mathbb{N}_1 (E,B) \equiv \frac{1}{2}  (\partial_x M_{NL} + \partial_t \widehat{\mu} P_{NL})  \label{nlPi}
\end{equation}
{\it That's the  system of equation of wave interaction of left in right wave due to arbitrary nonlinearity with the general temporal dispersion account. It is the principal result of this paper}.

\section{General equations of wave propagation in a 
metamaterial that is described by the lossless Drude model}

\subsection{Model for dispersion}
   The Drude model is a limit case of classical Lorentz model and represents a situation of a main contribution of free electrons, that explains its use in elementary conductivity theory, plazma physic and metamaterials. For this case we use relations from \cite{Wen}, accounting (\ref{mode(w)}):
\begin{equation}
    \varepsilon (\omega) = \left( 1 - \frac{\omega^2_{pe}}{\omega^2} \right) , \label{exp:e(w)}
\end{equation}
\begin{equation}
    \mu (\omega) =  \left( 1 - \frac{\omega^2_{pm}}{\omega^2} \right). \label{exp:mu(w)}
\end{equation}
This model is used by many authors \cite{Ziol3,Zhao}  et.  al., to describe the material properties of metamaterial.  
Energy density \eqref{energy} is positive at $\omega$ range for which (\ref{exp:e(w)}, \ref{exp:mu(w)}) are valid:
\begin{equation}
\begin{aligned}
W = \frac{d (\omega \varepsilon (\omega) )}{d \omega} E^2 + \frac{ d (\omega  \mu (\omega)) }{d \omega} H^2 =  \left( 1 + \frac{\omega^2_{pm}}{\omega^2} \right) E^2 + \left(  \frac{\omega^2_{pm}}{\omega^2} + 1 \right) H^2 > 0
\end{aligned}
\end{equation}
where $\omega_{pe}$ and $\omega_{pm}$ - parameters, dependent on the density, charge, and mass of the charge carrier.  These parameters are commonly known as the electric and magnetic plasma frequencies \cite{Wen}. The kernel $a(\omega)$ of the operator $\widehat{a}$ is:
\begin{equation}
 a (\omega)  = c^{-1} \sqrt{ \left( 1 - \frac{\omega^2_{pe}}{\omega^2} \right) \left( 1 - \frac{\omega^2_{pm}}{\omega^2} \right)}.
\end{equation}
After expansion $a(\omega)$ in Taylor series  in conditions of  $\omega << \omega_{pe}, \omega_{pm} $,  in a vicinity of $\omega=0$, we get:\\
\begin{equation}
\begin{aligned}
 \widehat{a} \eta (t) \approx c^{-1} \left[ \omega_{pe} \omega_{pm} \partial^{-2}_t - \frac{1}{2} \frac{\omega_{pe}^2+\omega_{pm}^2}{\omega_{pe} \omega_{pm}} + \left( \frac{1}{2 \omega_{pe} \omega_{pm}} + \frac{1}{8} \frac{(- \omega_{pe}^2 - \omega_{pm}^2)^2}{\omega_{pe}^3 \omega_{pm}^3}  \right) \partial^2_t \right] \eta (t). \label{P12-fin}
\end{aligned}
\end{equation}
The operator $\partial^{-1}_\alpha$ is defined as the integral:
\begin{equation}
\partial^{-1}_\alpha f(\alpha) = \int\limits_{0}^\alpha f(\beta) d\beta,
\end{equation}
As it is seen from the numerical analysis  in the range of frequencies $\omega < 0.5\omega_{pe}$, the first term is the leading one within the $0.005\%$  relative error.
We see also, the relative error for frequencies till $0.9 \omega_{pe}$ is less than $10\%$. \\
In the case of $\omega_{pm} = \omega_{pe}$ we find:
\begin{equation}
c a (\omega)  =  \sqrt{ \left( 1 - \frac{\omega^2_{pe}}{\omega^2} \right) \left( 1 - \frac{\omega^2_{pe}}{\omega^2} \right)} =  \sqrt{ \left( 1 - \frac{\omega^2_{pe}}{\omega^2} \right)^2} =  \left( 1 - \frac{\omega^2_{pe}}{\omega^2} \right),
\end{equation}
that already have algebraic form.

 Accounting all estimations, we left the only term in the relation \eqref{P12-fin}. 
Next, for compactness, we mark $\omega_{pe}$ as $p$, and $\omega_{pm}$ as $q$.
Plugging this minimal version of  \eqref{P12-fin} in the system \eqref{syslin} we obtain:
\begin{equation}
 \begin{aligned} \label{eq:sys13Ps}
 \partial_{x} \Pi    =  - c^{-1 } p q \partial^{-1}_t  \Pi  ,  \\
 \partial_{x} \Lambda    =  c^{-1 }  p q \partial^{-1}_t   \Lambda.
 \end{aligned}
 \end{equation}
 
Differentiating this system on $t$ once more, we write the resulting system, in which the right and left wave amplitudes are completely separated  
\begin{equation}
 \begin{aligned} \label{eq:sys13Ps}
 \partial_{xt} \Pi    =  - c^{-1 } p q   \Pi  ,  \\
 \partial_{xt} \Lambda    =    c^{-1 } p q  \Lambda .
 \end{aligned}
 \end{equation} 
Both equations describe the wave dispersion, ones are equivalent to 1+1 Klein-Gordon-Fock equation $\square\phi_{\pm}=m_{\pm}\phi_{\pm} $ with the mass parameter  $m_{\pm} =\pm c^{-1 } p q. $

\section{Kerr nonlinearity account for lossless Drude metamaterials}

\subsection{Equations of interaction of left and right waves with Kerr effect}

For nonlinear Kerr materials \cite{Christos, Christos2}, the third-order nonlinear part of polarisation \cite{Boyd,KuLe} has the form:
$$P_{NL} = \chi^{(3)} E^3$$
From \eqref{nlPi} we find $\mathbb{N}_1$:
\begin{equation}
\begin{aligned}
\mathbb{N}_1 \equiv \frac{1}{2} \widehat{\mu} ( \widehat{\mu}^{-1}\partial_x M_{NL} + \partial_t P_{NL}) = \frac{\chi^{(3)}}{2} \widehat{\mu} \partial_t E^3, \\
\end{aligned}
\end{equation}
Operator $\widehat{\mu}$ for the chosen model is just $\mu_0 (1- q^2 \partial_t^{-2})$. Moreover, effect of negative permeability was demonstrated at THz range \cite{Ipatov}. Hence $q^2 \partial_t^{-2} $ contribution prevails. Then, from  \eqref{nlsys} we obtain:
\begin{equation}
 \begin{aligned}  \label{eq:nldrude}
c \partial_{x} \Pi    -  p q \partial^{-1}_t    \Pi =  - \frac{\chi^{(3)}}{2} \mu_0 q^2 \partial^{-1}_t [\widehat{a}^{-1} (\Pi - \Lambda)]^3 ,  \\
 c \partial_{x} \Lambda    + p q \partial^{-1}_t   \Lambda =  \frac{\chi^{(3)}}{2} \mu_0 q^2 \partial^{-1}_t [\widehat{a}^{-1} (\Pi-\Lambda)]^3 ,
 \end{aligned}
 \end{equation}
The same approximation for the operator $\widehat{a}^{-1}$ reads as:
\begin{equation}
\widehat{a}^{-1} \eta (x,t) \approx  \frac{c}{pq} \partial_t^2\eta(x,t). \label{expansiona-1}
 \end{equation}
We substitute it to the system \eqref{eq:nldrude} and differentiate, denoting derivatives by indices for more compactness:
\begin{equation}
 \begin{aligned} \label{eq:sysnldrude}
c   \Pi_{xt}    +p q   \Pi =   -\frac{\mu_0 \chi^{(3)}c^3}{2p^3q} \left[  (\Pi - \Lambda)_{tt}\right]^3  ,  \\
c  \Lambda_{xt}   -  p q  \Lambda = \frac{ \mu_0 \chi^{(3)}c^3}{2p^3q}     \left[    (\Pi - \Lambda)_{tt}\right]^3 ,
 \end{aligned}
 \end{equation}
We consider this system as the main result of our work. The equivalent system is obtained by triple differentiation both equations of the system with respect to time and rescaling  $\Pi_{tt}=\alpha\pi,\,\Lambda_{tt}=\alpha\lambda,\,x=\beta\zeta$ with the choice $\alpha=\sqrt{\frac{2p^4q^2}{\mu_0\chi^{3}c^3}},\beta=\frac{c}{pq}.$ Then
\begin{equation}
\begin{array} {c}
\pi_{\zeta t}+\pi=- [(\pi-\lambda)^3]_{tt},\\
\lambda_{\zeta t}-\lambda=[(\pi-\lambda)^3]_{tt},
\end{array}
 \end{equation}
with extra boundary conditions.

Consider unidirectional case of \eqref{eq:sysnldrude} with $\Lambda = 0$, that corresponds special initial conditions from (\ref{bcq}) : $ (k(t) - \widehat{a} j(t))=0$ and valid till the effect of the left wave generation would be noticeable.
\begin{equation}\label{uniPi}
c  \Pi_{xt}    +p q   \Pi =   -\frac{\mu_0 \chi^{(3)}c^3}{2p^3q} \left[  \Pi_{tt}\right]^3. 
 \end{equation}

\subsection{Stationary solution}

\subsubsection{Linear case}
We introduce change of variables
\begin{equation}\label{xxi}
x = \eta, \xi = x - vt
\end{equation}
$v$ has dimension of speed. Then:
\begin{equation}
 \partial_t  = - v \partial_\xi , \partial_x  = \partial_\eta +  \partial_\xi ,      \,
\partial_{xt}  =  - v[ \partial_{\eta \xi} +  \partial_\xi^2] , \,
\Pi  (x,t) \rightarrow \mathsf{R} (\eta, \xi), \Lambda  (x,t) \rightarrow \mathsf{L} (\eta, \xi)
\end{equation}

\begin{equation}
 \begin{aligned} \label{eq:sys13Ps}
- v  [\partial_{\eta \xi} +  \partial_\xi^2 ] \mathsf{R}    =  -  p q c^{-1 }   \mathsf{R} ,  \\
- v   [\partial_{\eta \xi} +  \partial_\xi^2 ]  \mathsf{L}    =     p q  c^{-1 }  \mathsf{L}.
 \end{aligned}
 \end{equation} 
Start with only $\mathsf{R} $ wave. We declare the independence of $\mathsf{R}$  on $\eta$ as   a definition of stationary state:
$$\partial_\eta  \mathsf{R} = 0,$$
then:
\begin{equation} 
 \begin{aligned}
c^{-1 } p q  \mathsf{R}   =  v \partial_\xi^2 \mathsf{R} ,\label{stateq}
 \end{aligned}
 \end{equation}

Dimension of l.h.s. is a dimension of $\mathsf{k} \omega$:
\begin{equation}  \label{def:k}
\mathsf{k} \omega = \frac{pq}{c} .
\end{equation}
Also we found $v$:
\begin{equation} \label{def:v}
 \begin{aligned}
 v = \frac{\omega}{\mathsf{k}}.
 \end{aligned}
 \end{equation}
Let's return to \eqref{stateq} and rewrite it with accounting (\ref{def:k}, \ref{def:v}):
\begin{equation}  
 \begin{aligned}
 \frac{\omega}{\mathsf{k}}\partial_\xi^2 \mathsf{R} =  \omega \mathsf{k}  \mathsf{R}   ,
 \end{aligned}
 \end{equation}
 
\begin{equation}
 \begin{aligned}\label{R}
 \partial_\xi^2 \mathsf{R}- \mathsf{k}^2   \mathsf{R} = 0  ,
 \end{aligned}
 \end{equation} 
The boundary problem we solve means the solution domain specified by $x>0,t>0.$
For a   decaying boundary regime for $v>0$ the solution is: 
\begin{equation} 
 \begin{aligned} 
\mathsf{R} = A  \exp\left(  \mathsf{k} (x - vt) \right) ,
 \end{aligned}
 \end{equation}  

For the L-wave the equation differs only by the sign from \eqref{R}:
\begin{equation} 
 \partial_\xi^2 \mathsf{L} + \mathsf{k}^2   \mathsf{L} = 0  ,
 \end{equation} 
that gives oscillating solution: 
\begin{equation} 
\mathsf{L} = B \sin\left(  \mathsf{k}   (x-vt) \right).
 \end{equation} 
As we see, the negative value for $\mu$ drastically  changes the character of propagation of the waves $\mathsf{R} $ and  $\mathsf{L}$, which definition is given by \eqref{def:La} \eqref{def:Pi}.
\subsubsection{Nonlinear case}
The equation \eqref{uniPi} after the transition to variables \eqref{xxi} and use of stationary condition
\begin{equation}\label{stPi}
c  \Pi_{\xi\xi}    +p q   \Pi =   -\frac{\mu_0 \chi^{(3)}c^3v^6}{2p^3q} \left[  \Pi_{\xi\xi}\right]^3. 
 \end{equation}
Solving by Cardano formula the cubic equation with respect to $\Pi_{\xi\xi}$, one arrive at rather complicated  noninear oscillator
\begin{equation}
 \Pi_{\xi\xi}=F(\Pi), 
 \end{equation} 
which expansion with respect to $\Pi$ yields equations that may be directly integrated.
\section{Discussion and conclusion}

Using projection operators approach we derived the general system of equations \eqref{main}, that  describes  interaction between opposite directed  waves propagating in 1D-metamaterial, with  Kerr  nonlinearity .  The system is specified for a  lossless Drude model as \eqref{eq:sysnldrude}. The results may be used in experiments that investigate amplitude dependence of reflected wave from a metamaterial layer. 

Let us compare our results with Kinsler's method of derivation of a wave equation for left and right waves in time domain and scalar form \cite{Kin}:
\begin{equation}
\begin{aligned}
\partial_{x}  \Pi(x,t)   = \partial_t \alpha_r  \beta_r  \Pi + \frac{1}{2} \alpha_c \beta_r (\Pi - \Lambda), \label{kinsler}
\end{aligned}
\end{equation}
where $\alpha, \beta$ -  from expanding $\varepsilon$ and $\mu$:
\begin{equation}
  \begin{aligned}
  \varepsilon (\omega) = \varepsilon_r (\omega) + \varepsilon_c (\omega) = \alpha_r^2  (\omega) + \alpha_r (\omega) \alpha_c (\omega), \\
  \mu (\omega) = \mu_r (\omega) + \mu_c (\omega) = \beta_r^2  (\omega) + \beta_r (\omega) \beta_c (\omega)
  \end{aligned}
  \end{equation}
Index $c$ is called  the correction parameter, which "represents the discrepancy between the true values and the reference" \cite {Kin}. Correction parameters are depended only from nonlinearity. Index $r$ indicate "reference" values, close to the true medium properties, typically by including all the dispersive properties.  
As we see, projection operators method introduce left and right waves in more transparent manner (see also \eqref{kinsler}).

An investigation and solution of the obtained equations is planned in the nearest future.

\end{document}